\documentclass[runningheads]{ola}

\begin{document}

\pagestyle{headings}

\mainmatter

\title{Challenges in Automatic Software Optimization: the Energy Efficiency Case}

\titlerunning{Automatic Energy Optimization of Software}

\author{T. Fischbach\inst{1} \and E. Kieffer\inst{1} \and P. Bouvry\inst{1}}

\authorrunning{T. Fischbach, E. Kieffer and P. Bouvry}

\institute{University of Luxembourg\inst{1}, \\
\email{\{tobias.fischbach, emmanuel.kieffer, pascal.bouvry\}@uni.lu}
}

\maketitle
\vspace*{-20pt}
\section{Introduction}
\vspace*{-8pt}
With the advent of the Exascale capability allowing supercomputers to perform at least $10^{18}$ IEEE 754 Double Precision (64 bits) operations per second, many concerns have been raised regarding the energy consumption of high-performance computing code. Recently, Frontier operated by the Oak Ridge National Laboratory, has become the first supercomputer to break the exascale barrier \cite{Frontier2022}. In total, \emph{Frontier} contains 9,408 CPUs, 37,632 GPUs, and 8,730,112 cores. This world-leading supercomputer consumes about 21 megawatts which is truly remarkable as \emph{Frontier} was also ranked first on the Green500 list before being recently replaced. The previous top Green500 machine, MN-3 in Japan, provided 39.38 gigaflops per watt, while the \emph{Frontier} delivered 62.68 gigaflops per watt.
All these infrastructure and hardware improvements are just the tip of the Iceberg. Energy-aware code is now required to minimize the energy consumption of distributed and/or multi-threaded software.
For example, the data movement bottleneck is responsible for $35-60\%$ of a system's energy consumption during intra-node communication. In an HPC environment, additional energy is consumed through inter-node communication. This position paper aims to introduce future research directions to enter now in the age of energy-aware software.
The paper is organized as follows. First, we introduce related works regarding measurement and energy optimization. Then we propose to focus on the two different levels of granularity in energy optimization. 

\vspace*{-10pt}
\section{Related works }
\vspace*{-8pt}
\subsection{Measurement}
\vspace*{-8pt}
Optimizing software for energy efficiency necessitates thoroughly comprehending the energy consumption of different components (e.g., CPU, DRAM, GPU) at the process level. 

Commonly utilized tools, such as IPMI, are either too granular or lack the necessary precision, such as LIKWID and Slurm, for fine-grained measurements, as they are based on the same underlying technology RAPL \cite{noauthor_intel_nodate} (a low-level power capping framework that reports fine-grained CPU and RAM energy consumption). Haswell is the last generation that included on-die circuitry for measuring the consumed energy; current CPU generations only estimate the energy consumption. Based on validation experiments, it is necessary to adjust the reported energy consumption by an offset, but the reported values still show large deviations based on utilization as well as a large spread \cite{desrochers2016ValidationDRAMRAPL}\cite{lopez-novoa_exploring_2019}. In addition, AMD's implementation for Ryzen and Epyc lacks certain features (e.g., DRAM energy).

To enhance the accuracy of the energy consumption measurement without altering the motherboard, we propose an experimental setup based on sensible Hall sensors. The Hall sensors are placed strategically to monitor the energy consumption of different CPU packages and socket interconnections, DRAM, and PCIe connections. This approach accurately measures energy consumption without measuring overhead on the experiment itself. Among the challenges that are yet to be solved include the creation of reproducible software (e.g., interfering kernel tasks and binary reproducibility) and hardware environment (e.g., power states and frequency control) using NixOS to reduce noise. Due to the last generation with on-die hardware for power measurement, our results will be valided for Haswell-era CPUs.

\vspace*{-8pt}
\subsection{Automatic software optimization}
\vspace*{-8pt}
Up to now, the efforts toward an energy-aware compiler are mainly focused on either single-core or parallel workloads based on MPI. Embecosm employs super-optimization and machine learning to optimize LLVM-IR to reduce the energy consumption of software \cite{pallister_identifying_2015}. Other methods try to find the most energy-efficient combination of LLVM passes, a time-consuming and non-generalizable approach. Despite many efforts \cite{heinrich2017PredictingEnergyConsumptionMPI}, no precise model bridges the gap between coarse and fine-grained and accurately predicts the consumed energy.


%
%
%
%
%
%
\vspace*{-8pt}
\section{High-level Energy optimization }
\vspace*{-8pt}
\subsection{Surrogate-based algorithms}
Automatic software optimization is an NP-hard optimization problem that can only be tackled by means of heuristics or metaheuristics(see \cite{Varrette2020}) to obtain "good enough" approximations. The classical strategies consisting in applying genetic algorithms or, more generally, population-based metaheuristics are likely to fail since code evaluation is time-consuming. This explains why most of the academic research focused only on minimal benchmarks (e.g., matrix multiplication with LINPACK \cite{DDongarraLP03}). In automatic software optimization, the objective function has no explicit representation, and software code have to be simulated in a reproducible environment to get accurate and meaningful results. This is even truer when considering distributed software code on high-performance computing platforms. Therefore, we propose to consider surrogate-based optimization 
to minimize the number of simulations needed to evaluate the optimised code. Surrogate optimization attempts to find a global minimum of an objective function using a few objective function evaluations. For this purpose, the algorithm tries to balance the optimization process between two goals the search for a global minimum and the speed to obtain a good solution with few objective function evaluations. 
\vspace*{-8pt}
\subsection{Inferring decision rules}
\vspace*{-8pt}
Our decision to choose surrogate-based optimizers will definitely help us to consider larger problems, but we doubt that there is a significant advantage to applying a complex optimization framework for every chunk of code that may exist. Why can't we discover a set of rules connecting code patterns with code transformation? We hope to conduct this research to the point where we can automatically identify similarities between high-energy profiles. We could try to build machine learning models connecting LLVM analysis passes to LLVM transformation passes to infer common sense rules improving software energy consumption and change the pass sequence accordingly.
\vspace*{-10pt}
\section{Low-level Energy optimization}
\vspace*{-10pt}
The LLVM compiler framework \cite{lattner_llvm_2004} translates, transforms, and optimises source code using an intermediate representation (IR) in human-readable assembly\cite{grech_static_2015}.
From the results obtained during the high-level energy optimization, regions that consumed the most energy are identified and prioritized to create an energy-aware compiler.
\vspace*{-10pt}
\subsection{Energy-aware compiler}
\vspace*{-8pt}
A heterogeneous set of benchmarks is selected as optimization targets to reflect HPC's highly parallel and diverse nature. Finally, the energy consumption of performance and energy-optimised software is compared to identify the LLVM passes that reduce the energy requirements.
The LLVM-IR is inspected concerning code regions that can be used to reduce data movement, increase cache hits and prioritize energy-efficient instructions. Before enhancing instructions, candidates must be identified, and suitable replacements must be selected. Creating metrics correlating the measured energy consumption with IR code and the corresponding instructions should aid us in this process. While we intend to infer rules based on a consistent energy consumption model, this challenging task results in the large degree of freedom and non-linearity of interactions between instructions, hardware implementation, and compiler optimization.
Several critical factors related to communication are to be considered for optimization. The communication is considered by inter- and intra-node communication and cache misses \cite{kravets2022SoftwareCacheOptimizationBased}. The existing MPI implementations \cite{valter2022EnergyEfficiencyEvaluationOpenMP,venkatesh2015CaseApplicationobliviousEnergyefficient} are commonly used to implement intra- and inter-node communication effectively. On an intra-node communication scale, the energy consumption of communication is driven by inter-socket communication, core-to-core communication, and accelerator communication. The energy consumption of inter-node communication is affected by data movement \cite{gobieski_riptide_2022} and synchronization.

\vspace*{-8pt}
\subsection{Energy-aware language: a perspective}
\vspace*{-8pt}
The two proposed methods represent a "good enough" approach and try to enhance energy efficiency with languages that prioritize performance. We propose purpose-built languages that focus on energy efficiency instead. Expressing algorithms in more coarse language allows the usage of grammar evolution to automatic software creation. After sampling a sufficiently large state space, our proposed methods can optimize the vocabulary and grammar for energy efficiency. This purpose-built language can be extended to support more algorithms over time. Communication-heavy algorithms, like molecular dynamics simulations or deep learning, are ideal candidates due to their popularity and high-performance requirements.

\vspace*{-8pt}
\section{Conclusion and perspectives}
\vspace*{-10pt}
Improving the energy efficiency of software and algorithms is necessary for the era of exascale computing and IoT devices. However, current energy measurements are too coarse, especially for communication level and instruction level optimizations. This paper proposes a precise experiment based on sensitive hall sensors to measure the energy consumption of computation. We propose high-level energy optimizations based on surrogate algorithms to infer decision rules if and when LLVM performs specific transformations. Based on the high-level results, the most energy-consuming computation is identified, and low-level energy optimizations concerning intra- and inter-node communication are performed on an instruction level. Nevertheless, those methods try to transform performance-optimised languages to be more energy efficient. Lastly, we propose a purpose-built language with energy efficiency in mind.











\vspace*{-8pt}
\section*{Acknowledgement}
\vspace*{-10pt}
Tobias Fischbach acknowledges financial support of the Institute for Advanced Studies of the University of Luxembourg through a YOUNG ACADEMICS Grant (YOUNG ACADEMICS-2022-NETCOM).
\bibliographystyle{plain}
\vspace*{-15pt}
\bibliography{ola18}
\end{document}